\documentclass[10pt,a4paper,english]{article}
\usepackage[T1]{fontenc}
\usepackage[latin9]{inputenc}
\usepackage{color}
\usepackage{amsmath}
\usepackage{amssymb}
\usepackage{graphicx}
\makeatletter
\usepackage{amsfonts}
\usepackage{bbm}
\usepackage[T1]{fontenc}
\usepackage[latin9]{inputenc}
\usepackage{color}
\usepackage{amsmath}
\usepackage{amssymb}
\usepackage{graphicx}
\makeatletter
\usepackage{amsfonts}
\usepackage{bbm}
\usepackage{subfig}

\newcommand{\eps}{\varepsilon}
\newcommand{\bra}[1]{\langle#1|}
\newcommand{\ket}[1]{|#1\rangle}
\newcommand{\cc}{\mathbb{C}}
\newcommand{\tr}{\mathrm{Tr}}
\newcommand{\hh}{\mathcal{H}}
\newcommand{\bb}{\mathcal{B}}

\renewcommand{\>}{\rangle}
\newcommand{\rot}{\mathcal{R}}

\newcommand{\ot}[0]{\otimes}
\newcommand{\psii}[1]{\ket{\psi_{#1}}}
\newcommand{\fif}[1]{\ket{\phi_{#1}}}
\newcommand{\1}{\ket{1}}
\newcommand{\2}{\ket{2}}
\newcommand{\3}{\ket{3}}
\newcommand{\4}{\ket{4}}
\def\oper{{\mathchoice{\rm 1\mskip-4mu l}{\rm 1\mskip-4mu l}{\rm 1\mskip-4.5mu l}{\rm 1\mskip-5mu l}}}

\newtheorem{theorem}{Theorem}

\newtheorem{proposition}{Proposition}
\newtheorem{example}{Example}

\newcommand{\beq}{\begin{equation}}
\newcommand{\eeq}{\end{equation}}

\begin{document}
\title{\textbf{Geometry of entanglement witnesses \\ 
parameterized by $SO(3)$ group}}

\author{Dariusz Chru\'{s}ci\'{n}ski and Filip A. Wudarski\\
 Institute of Physics, Nicolaus Copernicus University,\\
 Grudzi\c{a}dzka 5/7, 87--100 Toru\'{n}, Poland}

\maketitle

\begin{abstract}
We characterize a set of positive maps in matrix algebra of $4\times4$ complex matrices. Equivalently,
we provide  a subset of entanglement witnesses in $\cc^4\otimes\cc^4$ parameterized by the rotation group $SO(3)$. Interestingly,
these maps/witnesses define two intersecting convex cones in the 3-dimensional parameter space. The existence of two  cones is related to the
topological structure of the underlying orthogonal group. We perform detailed analysis of the corresponding geometric structure.
\end{abstract}

\section{Introduction}

One of the most important problems of quantum information theory
is the characterization of mixed states of composed quantum systems \cite{hor,guhne}. In
particular it is of primary importance to test whether a given quantum state
exhibits quantum correlation, i.e. whether it is separable or entangled.

The most general method of solving separability problem is the one based on the notion of positive maps or equivalently
 entanglement witnesses (EWs). A state $\rho$ in $\mathcal{H}_A \ot \mathcal{H}_B$ is separable iff $(\oper_A \ot \Phi)\rho \geq 0$ for all
linear positive maps $\Phi : \mathcal{B}(\mathcal{H}_B) \rightarrow \mathcal{B}(\mathcal{H}_A)$. Recall, that
 a hermitian operator $W\in\bb(\hh_A\otimes\hh_B)$ is an entanglement witness \cite{terhal, horo2} iff:
i) it is not positively defined, i.e. $W \ngeqslant 0$, and ii) $\tr(W\sigma)\ge0$ for all separable states $\sigma$. Furthermore,
a bipartite state $\rho$ living in $\hh_A\otimes\hh_B$ is entangled iff there exists an EW $W$ detecting this state, i.e.
such that $\tr(W\rho)<0$. Due to the well known duality between maps and linear operators in $\bb(\hh_A\otimes\hh_B)$ these two
approaches are fully equivalent. Unfortunately, in spite of the considerable
effort (see e.g. \cite{maps-1}--\cite{maps-last}), the structure of positive maps/entanglement witnesses is rather poorly understood.

In this paper we analyze a class of positive maps  $\Phi : M_4(\cc) \rightarrow M_4(\cc)$ [$M_n(\cc)$ stands for an algebra
$n\times n$ complex matrices] parameterized by the rotation group $SO(3)$. This analysis extends our previous paper \cite{FilipI} where we discussed a class of maps
$\Phi : M_3(\cc) \rightarrow M_3(\cc)$ parameterized by the commutative group $SO(2)$. Our analysis shows that maps parameterized by $SO(3)$
belong to two intersecting coaxial cones. We  analyze the  geometric structure of these convex. Interestingly, our construction recovers
well known positive maps in $M_4(\cc)$: reduction  map and generalized Choi maps. We provide necessary and sufficient conditions for positivity and
perform detailed analysis of (in)decomposability. Our discussion is illustrated by several geometric figures.

It is hoped that our analysis sheds new light into the intricate structure of the convex cone of positive maps in matrix algebras.


\section{A class of positive maps in $M_n(\cc)$}

Let us recall a construction of a class of positive maps in $M_n(\cc)$ introduced by Kossakowski in \cite{koss} (for a slightly more general construction
cf. \cite{elsa}).  Let $f_\alpha$ $(\alpha=0,1,\ldots,n^2-1)$ denotes an orthonormal basis  in $M_n(\cc)$, such that $f_0=\frac{1}{\sqrt{n}} \mathbb{I}_n$, $f_k^*=f_k$, and
\beq
\tr(f_\alpha f_\beta)=\delta_{\alpha\beta},\quad \alpha,\beta=0,1,\ldots,n^2-1\ .
\eeq
Note, that $f_k$ ($k=1,2,\ldots,n^2-1)$ are traceless, i.e. $\tr f_k=0$ . Now the positive unital map is defined as follows \cite{koss}
\beq\label{map}
\Phi_R(X)=\frac{1}{n}\mathbb{I}_n\tr\,X+\frac{1}{n-1}\sum_{\alpha,\beta=1}^{n^2-1}f_\alpha R_{\alpha\beta}\tr(f_\beta X)\ ,
\eeq
where $R_{kl}$ is an arbitrary rotation matrix from $O(n^2-1)$. Note, that a dual map $\Phi_R^\#$ defined by
$$   {\rm Tr}[ Y \Phi_R(X)] =: {\rm Tr}[\Phi^\#_R(Y) \cdot X] \ , $$
reads
\beq\label{map}
\Phi_R^\#(Y)=\frac{1}{n}\mathbb{I}_n\tr\,Y+\frac{1}{n-1}\sum_{\alpha,\beta=1}^{n^2-1}f_\alpha R_{\beta\alpha}\tr(f_\beta Y)\ ,
\eeq
and hence
\begin{equation}\label{}
    \Phi_R(\mathbb{I}_n) = \mathbb{I}_n\ , \ \ \ \  \Phi^\#_R(\mathbb{I}_n) = \mathbb{I}_n\ .
\end{equation}
Note, that if $R$ corresponds to reflection in $\mathbb{R}^{n^2-1}$, i.e. $R_{\alpha\beta} = - \delta_{\alpha\beta}$, one easily finds
\begin{equation}\label{}
    \Phi_R(X) = \frac{1}{n-1} ( \mathbb{I}_n\tr\,X - X)\ ,
\end{equation}
and hence it reproduced the reduction map in $M_n(\cc)$.

Consider now a special class of maps $\Phi_R$ corresponding to
\begin{equation}\label{}
    R = \mathcal{R} \oplus (-\mathbb{I}_{n(n-1)})\ ,
\end{equation}
where $\mathcal{R} \in O(n-1)$. In particular if $\mathcal{R} = - \mathbb{I}_{n-1}$ one reproduces reduction map.

Let $\{f_k\}$ $(k=1,\ldots,n^2-1)$ denote generalized Gell-Mann matrices defined as follows: let
 $|1\>,\ldots,|n\>$ be an orthonormal basis in $\mathbb{C}^n$ and define
\begin{eqnarray}
d_l &=& \frac{1}{\sqrt{l(l+1)}}\Big(\sum_{k=1}^l \ket{k}\bra{k}-l \ket{l+1}\bra{l+1}\Big)\ ,
\end{eqnarray}
for $l=1,\ldots,n-1$, and
\begin{eqnarray}
u_{kl} &=&\frac{1}{\sqrt{2}}(\ket{k}\bra{l}+\ket{l}\bra{k}),\\
v_{kl} &=&\frac{-i}{\sqrt{2}}(\ket{k}\bra{l}-\ket{l}\bra{k}),
\end{eqnarray}
for $k<l$. It is clear that $n^2$ Hermitian matrices $(f_0,d_l,u_{kl},v_{kl})$ define a proper orthonormal basis in $M_n(\cc)$.
One easily finds
\begin{eqnarray}
  \Phi_R(|i\>\<i|) &=& \sum_{j=1}^n \Phi_{ij} |j\>\<j|\ , \\
   \Phi_R(|i\>\<j|) &=& - \frac{1}{n-1}\, |i\>\<j|\ , \ \ \ i\neq j\ ,
\end{eqnarray}
where the matrix $\Phi_{ij}$ reads as follows
\begin{equation}\label{}
    \Phi_{ij} = \frac 1n + \frac{1}{n-1} \sum_{k,l=1}^{n-1} \, \<j|d_l|j\>\, \mathcal{R}_{kl}\, \<i|d_k|i\> \ .
\end{equation}
One shows \cite{OSID-07} that the matrix $\Phi_{ij}$ is doubly stochastic.

The corresponding entanglement witness is defined as follows
\begin{equation}\label{}
    W := n(n-1) (\oper \ot \Phi_R) P^+_n\ ,
\end{equation}
where $P^+_n$ denotes maximally entangled state (we add the factor `$n(n-1)$' to simplify the form of $W$).  One finds
\begin{equation}\label{Wa}
    W = \sum_{i,j=1}^n |i\>\<j| \ot W_{ij}\ ,
\end{equation}
where $W_{ij} = - |i\>\<j|$ for $i \neq j$, and
\begin{equation}\label{Wb}
    W_{ii} := (n-1) \Phi_R(|i\>\<i|) = (n-1) \sum_{j=1}^n \Phi_{ij} |j\>\<j|\ .
\end{equation}

\begin{example} If $n=3$ one obtains the following formula for the $3 \times 3 $ matrix $\,\Phi_{ij}$
\begin{equation}\label{Phi-3}
  \Phi_{ij}  =   \frac 12 \left(  \begin{array}{ccc} a & b & c \\ c & a &  b \\ b & c & a \end{array} \right)\ ,
\end{equation}
where $a,b,c \geq 0$ are parameterized by the $SO(2)$ rotation as follows
\begin{eqnarray}   \label{abc}
a = \frac{2}{3}\,(1+\cos\alpha)\ , \ \ \
b = \frac{2}{3}\left(1-\frac{1}{2}\cos\alpha-\frac{\sqrt{3}}{2}\sin\alpha\right) \ ,\ \ \
c  = \frac{2}{3}\left(1-\frac{1}{2}\cos\alpha+\frac{\sqrt{3}}{2}\sin\alpha\right)\ .
\end{eqnarray}
This class of maps was analyzed recently in \cite{FilipI}. Note, that $a+b+c=2$ and hence  formulae (\ref{abc}) define an ellipse on the $bc$-plane. Actually, this ellipse is defined by the following condition
\begin{equation}\label{ellipse}
    bc = (1-a)^2\ ,
\end{equation}
and  for $a\leq 1$ it belongs to the boundary of a convex set of entanglement  witnesses defined by the well known conditions \cite{Cho-Kye}
\begin{equation}\label{}
  i)\ \ 0 \leq a < 2\ ,\ \  \ \ \ ii)\ \ a+b+c \geq 2\ , \ \ \ \ \  iii)\ \ a \leq 1\ \Rightarrow \ bc \geq (1-a)^2\ .
\end{equation}
Moreover, it was shown \cite{kye-optimal,Gniewko-opt}
that for $a\leq 1$ our family defines optimal entanglement witnesses.
Interestingly, for $a < 1$ these witnesses are even exposed \cite{kye-exposed}.

\end{example}


\section{Entanglement witnesses in $\cc^4 \ot \cc^4$}

In this section we elaborate the construction of entanglement witnesses defined by (\ref{Wa})--(\ref{Wb}) for $n=4$.
One finds for the corresponding $\Phi_{ij}$ matrix
\begin{equation}\label{Phi-3}
  \Phi_{ij}  =   \frac 13 \left(  \begin{array}{cccc} a_1 & b_1 & c_1 & d_1 \\ d_2& a_2 & b_2 &  c_2 \\ c_3 & d_3 & a_3 & b_3 \\ b_4 & c_4 & d_4 & a_4 \end{array} \right)\ ,
\end{equation}
where $a,b,c,d \geq 0$ are parameterized by $\mathcal{R} \in SO(3)$. Any such $\mathcal{R}$ may be represented as follows
\begin{equation}\label{eul}
\rot=\left(\begin{array}{ccc}c_\alpha c_\gamma-c_\beta s_\alpha s_\gamma & c_\gamma s_\alpha+c_\alpha c_\beta s_\gamma & s_\beta s_\gamma \\
-c_\beta c_\gamma s_\alpha-c_\alpha s_\gamma & c_\alpha c_\beta c_\gamma -s_\alpha s_\gamma & c_\gamma s_\beta \\
s_\alpha s_\beta & -c_\alpha s_\beta & c_\beta\end{array}\right) \ ,
\end{equation}
where $(\alpha,\beta,\gamma)$ denote Euler angles and to simply notation we use $s_\alpha := \sin\alpha$ and $c_\alpha:= \cos\alpha$.
Unfortunately, the formulae for matrix elements $\Phi_{ij}$ are quite involved (see the Appendix).
Moreover, contrary to $n=3$ the doubly stochastic matrix $\Phi_{ij}$ is no longer circulant.
To simply our analysis we use the following simple observation: let $U_{kl}$ be a set of unitary matrices defined as follows
\begin{equation}\label{}
 U_{kl} =\sum_{m=1}^{4}\, \omega^{km}\ket{m}\bra{m\oplus l}\ ,
\end{equation}
where $\omega= e^{2\pi i/4}$ and $m\oplus l$ denotes addition modulo $4$. Let $P_{kl} =
(\mathbb{I}_4 \ot U_{kl}) P^+_4 (\mathbb{I}_4 \ot U_{kl}^\dagger)$ and define
\begin{equation}\label{}
    \widetilde{W} := \sum_{k,l} {\rm Tr}(W P_{kl})\, P_{kl} \ .
\end{equation}
It turns out \cite{Werner} that  if $W$ is an entanglement witness then  $\widetilde{W}$ is an entanglement witness as well.
One finds that $\widetilde{W}$ is again defined by formulae (\ref{Wa})--(\ref{Wb}) with $\Phi_{ij}$ given by the following circulant matrix
\begin{equation}\label{Phi-3-c}
  \Phi_{ij}  =   \frac 13 \left(  \begin{array}{cccc} a & b & c & d \\ d& a & b &  c \\ c & d & a & b \\ b & c & d & a \end{array} \right)\ ,
\end{equation}
where
 \beq
 a :=\frac{1}{4}\sum_{i=1}^4\, a_i\ ,\qquad  b:=\frac{1}{4}\sum_{i=1}^4\, b_i\ ,\qquad  c:=\frac{1}{4}\sum_{i=1}^4\, c_i\ ,\qquad    d:=\frac{1}{4}\sum_{i=1}^4\, d_i\ ,
 \eeq
read as follows
\begin{eqnarray}  \label{fa}
 a&=&\frac{1}{4}\Big[3+c_{\alpha+\gamma}(1+c_\beta)+c_\beta\Big],\\
b&=&\frac{1}{4}\Big[3+\frac{1}{6}(s_\alpha s_\gamma-c_\alpha c_\beta c_\gamma-3c_\alpha c_\gamma+3c_\beta s_\alpha s_\gamma-2c_\beta) +\frac{1}{2\sqrt{3}}(3c_\gamma s_\alpha+\nonumber\\
&+&3c_\alpha c_\beta s_\gamma+c_\beta c_\gamma s_\alpha+c_\alpha s_\gamma)+\frac{2}{3\sqrt{2}}s_\beta(2c_\gamma +c_\alpha)-\frac{2}{\sqrt{6}}s_\alpha s_\beta\Big],\\
c&=&\frac{1}{4}\Big[ 3-\frac{1}{3}(2c_\alpha c_\beta c_\gamma-2s_\alpha s_\gamma+c_\beta)-\frac{2}{3\sqrt{2}}s_\beta(c_\gamma-c_\alpha)+\frac{2}{\sqrt{6}} s_\beta (s_\gamma+s_\alpha)-\nonumber\\
&-&\frac{1}{\sqrt{3}}(c_\gamma s_\alpha+c_\alpha c_\beta s_\gamma-c_\beta c_\gamma s_\alpha -c_\alpha s_\gamma)\Big],\\
d&=&\frac{1}{4}\Big[3+\frac{1}{6}(s_\alpha s_\gamma-c_\alpha c_\beta c_\gamma-3c_\alpha c_\gamma+3c_\beta s_\alpha s_\gamma-2c_\beta)-\frac{1}{2\sqrt{3}}(3c_\beta c_\gamma s_\alpha+\nonumber\\
&+&3c_\alpha s_\gamma+c_\gamma s_\alpha+c_\alpha c_\beta s_\gamma)-\frac{2}{3\sqrt{2}}s_\beta(c_\gamma+2c_\alpha)-\frac{2}{\sqrt{6}}s_\gamma s_\beta \Big]\ . \label{fb}
\end{eqnarray}
Interestingly, the EW corresponding to the reduction map does not belong to this class since reflection is not a proper rotation from $SO(3)$. To include such case let us replace $\mathcal{R} \longrightarrow - \mathcal{R}$. It is clear that if $\mathcal{R}$ is a proper rotation from $SO(3)$ then $-\mathcal{R} \in O(3)$. Using the same arguments one obtains a new class of EWs $\widetilde{W}'$ with $(a,b,c,d)$ replaced by
\begin{eqnarray}
 a'&=&\frac{1}{4}\Big[3-c_{\alpha+\gamma}(1+c_\beta)-c_\beta\Big], \label{ffa}\\
b'&=&\frac{1}{4}\Big[3-\frac{1}{6}(s_\alpha s_\gamma-c_\alpha c_\beta c_\gamma-3c_\alpha c_\gamma+3c_\beta s_\alpha s_\gamma-2c_\beta) -\frac{1}{2\sqrt{3}}(3c_\gamma s_\alpha+\nonumber\\
&+&3c_\alpha c_\beta s_\gamma+c_\beta c_\gamma s_\alpha+c_\alpha s_\gamma)-\frac{2}{3\sqrt{2}}s_\beta(2c_\gamma +c_\alpha)+\frac{2}{\sqrt{6}}s_\alpha s_\beta\Big],\\
c'&=&\frac{1}{4}\Big[ 3+\frac{1}{3}(2c_\alpha c_\beta c_\gamma-2s_\alpha s_\gamma+c_\beta)+\frac{2}{3\sqrt{2}}s_\beta(c_\gamma-c_\alpha)-\frac{2}{\sqrt{6}} s_\beta (s_\gamma+s_\alpha)+\nonumber\\
&+&\frac{1}{\sqrt{3}}(c_\gamma s_\alpha+c_\alpha c_\beta s_\gamma-c_\beta c_\gamma s_\alpha -c_\alpha s_\gamma)\Big],\\
d'&=&\frac{1}{4}\Big[3-\frac{1}{6}(s_\alpha s_\gamma-c_\alpha c_\beta c_\gamma-3c_\alpha c_\gamma+3c_\beta s_\alpha s_\gamma-2c_\beta)+\frac{1}{2\sqrt{3}}(3c_\beta c_\gamma s_\alpha+\nonumber\\
&+&3c_\alpha s_\gamma+c_\gamma s_\alpha+c_\alpha c_\beta s_\gamma)+\frac{2}{3\sqrt{2}}s_\beta(c_\gamma+2c_\alpha)+\frac{2}{\sqrt{6}}s_\gamma s_\beta \Big]\ . \label{ffb}
\end{eqnarray}
Formulae for $(a,b,c,d)$ and $(a',b',c',d')$ provide an analog of much simpler relations (\ref{abc}) for $n=3$.

\section{Geometry of cones}

Now comes a natural question: what is the geometric representation of the above relations?
For $n=3$ formulae (\ref{abc}) give rise to an ellipse in the $bc$-plane (cf. \cite{FilipI}).
Interestingly for $n=4$ an elegant geometric picture arises as well.
Actually, it was the original motivation for this paper. Numerical analysis shows the following picture in $(b,c,d)$ coordinates
 (recall, that $a+b+c+d=3$): formulae  (\ref{fa})--(\ref{fb}) and (\ref{ffa})--(\ref{ffb}) give rise to two intersecting  cones (see Fig. (a))
\begin{figure}[htp]   \label{cones}
 \centering
\subfloat[]{\label{cones}
\includegraphics[scale=0.8]{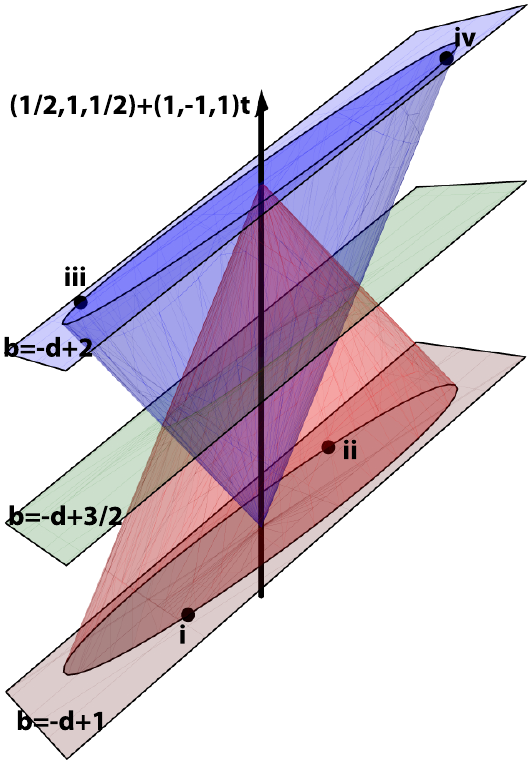}} 
  \caption{
 Cone I [blue] is bounded by a plane $b=-d+2$ and cone II [red] by a plane $b=-d+1$. Both cones are coaxial with axis
  described by the line $(\frac{1}{2}, 1,\frac{1}{2})+(1,-1,1)t$. Moreover, the special points are marked: associated with
  generalized Choi  maps (i), (ii), (iv)  and connected with reduction map (iii).}
\end{figure}
The above cones are described by the following equations: Cone I
\begin{eqnarray} \label{blue}
(b-2)^2+(2c-3)^2+(d-2)^2+4bc+4cd-2bd = 9 \ ,
\end{eqnarray}
and Cone II
\begin{eqnarray}  \label{red}
(b-1)^2+(2c-3)^2+(d-1)^2+4bc+4cd-2bd = 6\ ,
\end{eqnarray}
with the constraints that $-b+1\le d\le -b+2$. The vertices of these cones are located at
$b=\frac{1}{2},\ c=1,\ d=\frac{1}{2}$ and $b=1,\ c=\frac{1}{2},\ d=1$ for cones I and II, respectively.
They intersect along an ellipse in the plane $d=-b+\frac{3}{2}$.

Let us analyze the intersection of the Cone I defined by (\ref{blue}) with the plane $b+d=2$. One finds
\begin{equation}\label{}
    (d-1)^2 + (c-\frac 12)^2 = \frac 14 \ .
\end{equation}
Similarly the intersection of the Cone II defined by (\ref{red}) with the plane $b+d=1$ gives
\begin{equation}\label{}
    (c-1)^2 + (d-\frac 12)^2 = \frac 14 \ .
\end{equation}
Taking into account $a+b+c+d=3$ one finds
\begin{eqnarray}\label{red-a}
a&=&1-c, \nonumber\\
b&=& 1\pm\sqrt{c(1-c)},\\
d&=&1\mp\sqrt{c(1-c)}, \nonumber
\end{eqnarray}
with $c \in [0,1]$ for the ellipse I on $b+d=2$ plane
and
\begin{eqnarray}  \label{blue-a}
a &=&1\pm\sqrt{d(1-d)}, \nonumber \\
b &=& 1-d,\\
c &=& 1\mp\sqrt{d(1-d)}\nonumber
\end{eqnarray}
with $d \in [0,1]$ for the ellipse II on $b+d=1$ plane.  Interestingly, ellipse I [blue] satisfies
\begin{equation}\label{blue-1}
    a+c=1\ , \ \ \ b+d=2\ ,
\end{equation}
whereas the ellipse II [red] satisfies
\begin{equation}\label{red-1}
    a+c=2\ , \ \ \ b+d=1\ .
\end{equation}
Formulae (\ref{blue-a}) imply
\begin{equation}\label{}
    bd = (1-a)^2\ .
\end{equation}
Similarly, formulae (\ref{red-a}) imply
\begin{equation}\label{}
    ac = (1-b)^2\ .
\end{equation}
These two equations provide an analog of the well known condition (\ref{ellipse}) for $n=3$.

Let us observe that these two ellipses  contain already known positive maps:

\begin{itemize}
\item $\Phi[0,1,1,1]\,$  -- reduction map  [point (iii) in Fig. 1],
\item $\Phi[1,1,1,0]\,$  -- generalized Choi map  [point (i) in Fig. 1],
\item $\Phi[1,0,1,1]\,$ -- generalized Choi map  [point (ii) in Fig. 1],
\item $\Phi[1,1,0,1]\,$ --  [point (iv) in Fig. 1].
\end{itemize}
Note, however, that another generalized Choi map $\Phi[2,1,0,0]$ does not belong to our class.


\section{(In)decomposability}

In this section we analyze the issue of indecomposability of $\Phi[a,b,c,d]$.

\begin{theorem}
$\Phi[a,b,c,d]$ is decomposable if and only if $b=d$.
\end{theorem}
{\bf Proof}:  let us consider a state given by (unnormalized) density matrix
\begin{equation}\label{}
    \rho_\eps =   \sum_{i=1}^4 \Big[ |ii\>\<ii| + \eps |i,i+1\>\<i,i+1| + |i,i+2\>\<i,i+2| + \eps^{-1} |i,i+3\>\<i,i+3| \Big] + \sum_{i\neq j} |ii\>\<jj|  \ ,
\end{equation}
with $\eps > 0$. One easily checks that $\rho_\eps$ is PPT. One finds
\beq
\tr(W[a,b,c,d] \rho_\eps)=-12+4(a+c)+4\eps^{-1}d+4\eps b=4[d\eps^{-1}+b\eps-(b+d)] = 4\eps^{-1}[ b \eps^2 - (b+d)\eps + d]\ ,
\eeq
and hence $\tr(W[a,b,c,d] \rho_\eps) < 0$ iff there exists $\eps >0$ such that $b \eps^2 - (b+d)\eps + d < 0$. The corresponding discriminant reads
$$ \Delta = (b+d)^2 - 4bd = (b-d)^2 \ , $$
and hence $b \eps^2 - (b+d)\eps + d < 0$ if $\eps \in (\eps_-,\eps_+)$ with
\begin{equation*}\label{}
    \eps_\pm = \frac{b+d \pm |b-d|}{2b}\ .
\end{equation*}
Note, that $\eps_+ > \eps_-$ if and only if $b\neq d$. This way we have proved that if $b \neq d$ then $\Phi[a,b,c,d]$ is indecomposable.
Now we show that if $b=d$, then  $\Phi[a,b,c,d]$ is decomposable.
We find the corresponding decomposition
\beq
W[a,b,c,d]=P[a,b,c,d]+Q[a,b,c,d]^\Gamma,
\eeq
where, $P[a,b,c,d]$ and $Q[a,b,c,d]$ are positive operators. One has
\begin{equation}\label{}
  P[a,b,c,d]   =  \sum_{i=1}^4 \Big[ a |ii\>\<ii|  - (1-b)[\, |ii\>\<i+1,i+1| + |ii\>\<i+3,i+3|  \,]
  - (1-c)|ii\>\<i+2,i+2| \Big]\ ,
\end{equation}
and
\begin{eqnarray}\label{}
  Q[a,b,c,d]   &=&  \sum_{i=1}^4 \Big[ b|i,i+1\>\<i,i+1| + c|i,i+2\>\<i,i+2| + b|i,i+3\>\<i,i+3|  \nonumber \\ & -&
   b(\, |i,i+1\>\<i+1,i| + |i,i+3\>\<i+3,i|  \,)
  - c|i,i+2\>\<i+2,i| \Big]\ .
\end{eqnarray}
It is clear that $Q[a,b,c,d] \geq 0$. Now, to prove that $P[a,b,c,d] \geq 0$ one needs to show that the following circulant matrix
\begin{equation}\label{}
    A = \left( \begin{array}{cccc} a & b-1 & c-1 & b-1 \\ b-1 & a & b-1 & c-1 \\ c-1 & b-1 & a & b-1 \\ b-1 & c-1 & b-1 & a \end{array} \right) \ ,
\end{equation}
is positive. One finds for the eigenvalues of $A$: $\{0,4(1-b),2(2-b-c),2(2-b-c)\}$. Note, that $b+d \leq 2$ and hence $b=d\leq 1$.
Moreover, $2-b-c = (a+c)-1 \geq 0$. Hence $A \geq 0$.

     \hfill $\Box$

\begin{figure}[htp]
\begin{center}
 \label{cones2}\includegraphics[scale=0.8]{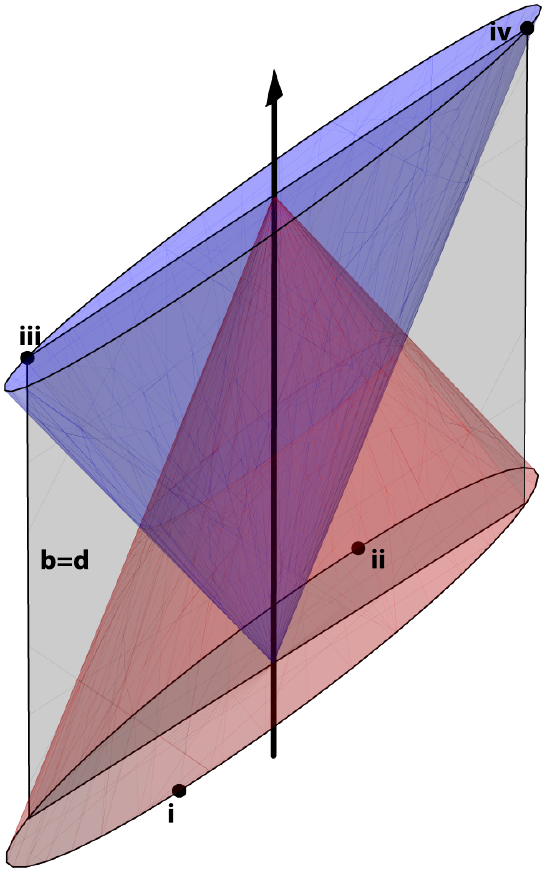}
  \caption{The intersection of the plane $b=d$ and  of the blue and red cones represents decomposable EWs.}
\end{center}
\end{figure}
%

\section{Structural Physical Approximation }

For any entanglement witness $W$  in $\hh_A\otimes\hh_B$  such that ${\rm Tr}\, W = 1$
one defines its structural physical approximation (SPA)
\beq
W(p)=(1-p)W+\frac{p}{d_A d_B}\mathbbm{1}_A\otimes\mathbbm{1}_B,
\eeq
with $p\ge p^*$, where $p^*$ is the smallest value of $p$ such that $W(p)\ge0$. Thus, SPA defines a legitimate
quantum state in $\hh_A\otimes\hh_B$. The conjecture of Korbicz et al. \cite{spa1} (see also \cite{spa2}) states that,
if $W$ is an optimal EW, then its SPA defines separable state. It was supported by several examples (see e.g. \cite{jp1}).

In a recent paper \cite{FilipI} it was conjectured that all entanglement witnesses $W[a,b,c]$ with $a,b,c$ satisfying
(\ref{ellipse}) and $a \leq 1$ are optimal. Actually, this conjecture was proved by Ha and Kye \cite{kye-optimal} (see also \cite{Gniewko-opt}).
It was shown \cite{FilipI} that $W[a,b,c]$ support SPA conjecture \cite{spa1}. Now we prove the following

\begin{proposition}
If $a,b,c,d$ satisfy (\ref{blue-a}) or (\ref{red-a}) then the structural physical approximation to $W[a,b,c,d]$ provides a
separable state in $\mathbb{C}^4 \ot \mathbb{C}^4$.
\end{proposition}
{\bf Proof}: Let us consider SPA for our class
\beq
\mathbf{W}(p)=(1-p)W+\frac{p}{16}\mathbb{I}_4\otimes\mathbb{I}_4\ .
\eeq
Now, $\mathbf{W}(p)\ge0$ for $p\ge p^*$, where the critical value $p^*$ is given by
\beq
p^*=\frac{4(a-3)}{3+4(a-3)}.
\eeq
It turns out that $W(p^*)$ can be represented as
\beq
\mathbf{W}(p^*)=\frac{1}{4[3+4(a-3)]}(\sigma_{12}+\sigma_{13}+\sigma_{14}+\sigma_{23}+\sigma_{24}+\sigma_{34}+\sigma_{d}),
\eeq
where
\beq
\sigma_{ij}=\ket{ij}\bra{ij}+\ket{ji}\bra{ji}+\ket{ii}\bra{ii}+\ket{jj}\bra{jj}-\ket{ii}\bra{jj}-\ket{jj}\bra{ii},
\eeq
and
\beq
\sigma_d=\sum_{i=1}^4\Big((2b+c+d-1)\ket{i,i+1}\bra{i,i+1}+(2c+b+d-1)\ket{i,i+2}\bra{i,i+2}+(2d+b+c-1)\ket{i,i+3}\bra{i,i+3}   \Big).
\eeq
Due to the fact that $\sigma_{ij}$  are PPT and supported on $\cc^2\otimes\cc^2$,  they are separable. Moreover, $\sigma_d$ is separable,
whenever it defines a legitimate quantum state, that is, when
\begin{eqnarray} \label{spa-3}
2b+c+d&\ge&1,\nonumber\\
2c+b+d&\ge&1,\\
2d+b+c&\ge &1.\nonumber
\end{eqnarray}
It is straightforward to show that both conditions (\ref{blue-a}) and (\ref{red-a}) imply (\ref{spa-3}) which ends the proof. \hfill $\Box$

Interestingly, the above three 2-dimensional planes:
$$ 2b+c+d=1\ , \ \ 2c+b+d=1\ , \ \  2d+b+c = 1 \ ,  $$
intersect at $b=c=d=1/4$.


\section{Conclusions}

We analyzed a class of positive maps in $M_4(\cc)$ (or equivalently class of EWs in $\bb(\cc^4\otimes\cc^4)$). This construction generalizes
analysis provided in \cite{FilipI} for $n=3$ by replacing commutative group $SO(2)$ by the noncommutative rotation group $SO(3)$.
 We formulated necessary and sufficient conditions for positivity of $\Phi[a,b,c,d]$ and described
the geometric structure of the convex set formed by these maps. Interestingly, there are two proper cones in the 3-dimensional space
parameterized by $(b,c,d)$. It was shown that for $b\neq d$ all maps are indecomposable and hence can be used to detect PPT entangled states. Moreover,
maps satisfying (\ref{spa-3})  support SPA conjecture \cite{spa1}. We provided two natural 1-parameter subclasses of maps (cf. formulae
(\ref{blue-a}) and (\ref{red-a})) -- two ellipses in the parameter space --
which are direct generalizations  of 1-parameter class of maps analyzed in \cite{FilipI}.

In a forthcoming paper we are going to analyze further properties of maps $\Phi[a,b,c,d]$. In particular optimality and atomicity.
Finally, it would be
interesting to provide the analogous construction for arbitrary $n$.

\section*{Appendix}

\begin{align*}
& a_1 = \frac{3}{4} + \frac{1}{2}R_{11}+\frac{1}{2\sqrt{3}}R_{12}+\frac{1}{2\sqrt{6}}R_{13}+ \frac{1}{2\sqrt{3}}R_{21}+\frac{1}{6}R_{22}+\frac{1}{6\sqrt{2}}R_{23}+ \frac{1}{2\sqrt{6}}R_{31}+\frac{1}{6\sqrt{2}}R_{32}+\frac{1}{12}R_{33}\  ,\\
& a_2= \frac{3}{4} -\frac{1}{2\sqrt{3}}R_{21}+\frac{1}{6}R_{22}+\frac{1}{6\sqrt{3}}R_{23}+ \frac{1}{2}R_{11}-\frac{1}{2\sqrt{3}}R_{12}-\frac{1}{2\sqrt{6}}R_{13}- \frac{1}{2\sqrt{6}}R_{31}+\frac{1}{6\sqrt{2}}R_{32}+\frac{1}{12}R_{33}\ ,\\
& a_3= \frac{3}{4} -\frac{1}{3\sqrt{2}}R_{32}+\frac{1}{12}R_{33}+\frac{2}{3}R_{22}-\frac{1}{3\sqrt{2}}R_{23}\ ,\\
& a_4= \frac{3}{4}+ \frac{9}{12}R_{33}\ , \\
%
%
& b_1= \frac{3}{4} -\frac{1}{2}R_{11}+\frac{1}{2\sqrt{3}}R_{12}+\frac{1}{2\sqrt{6}}R_{13}- \frac{1}{2\sqrt{3}}R_{21}+\frac{1}{6}R_{22}+\frac{1}{6\sqrt{2}}R_{23}- \frac{1}{2\sqrt{6}}R_{31}+\frac{1}{6\sqrt{2}}R_{32}+\frac{1}{12}R_{33}\ ,\\
& b_2= \frac{3}{4}+ \frac{1}{\sqrt{3}}R_{12}-\frac{1}{2\sqrt{6}}R_{13}-\frac{1}{3}R_{22}+ \frac{1}{6\sqrt{2}}R_{23}-\frac{1}{3\sqrt{2}}R_{32}+\frac{1}{12}R_{33}\ ,\\
& b_3= \frac{3}{4} -\frac{3}{12}R_{33}+\frac{1}{\sqrt{2}}R_{23}\ ,\\
& b_4= \frac{3}{4} -\frac{3}{2\sqrt{6}}R_{31}-\frac{1}{2\sqrt{2}}R_{32}-\frac{3}{12}R_{33}\ ,\\
%
%
& c_1= \frac{3}{4} -\frac{1}{\sqrt{3}}R_{12}+\frac{1}{2\sqrt{6}}R_{13}-\frac{1}{3}R_{22}+ \frac{1}{6\sqrt{2}}R_{23}-\frac{1}{3\sqrt{2}}R_{32}+\frac{1}{12}R_{33}\ ,\\
& c_2= \frac{3}{4} -\frac{1}{2\sqrt{2}}R_{23}+\frac{3}{2\sqrt{6}}R_{13}-\frac{3}{12}R_{33}\ ,\\
& c_3= \frac{3}{4} + \frac{1}{2\sqrt{6}}R_{31}+\frac{1}{6\sqrt{2}}R_{32}+\frac{1}{12}R_{33}- \frac{1}{\sqrt{3}}R_{21}-\frac{1}{3}R_{22}-\frac{1}{3\sqrt{2}}R_{23}\ ,\\
& c_4= \frac{3}{4} + \frac{3}{2\sqrt{6}}R_{31}-\frac{1}{2\sqrt{2}}R_{32}-\frac{3}{12}R_{33}\ ,\\
%
& d_1= \frac{3}{4} -\frac{3}{2\sqrt{6}}R_{13}-\frac{1}{2\sqrt{2}}R_{23}-\frac{1}{4}R_{33}\ ,\\
& d_2 = \frac{3}{4} + \frac{1}{2\sqrt{3}}R_{21}+\frac{1}{6}R_{22}+\frac{1}{6\sqrt{2}}R_{23}-\frac{1}{2}R_{11}- \frac{1}{2\sqrt{3}}R_{12}-\frac{1}{2\sqrt{6}}R_{13}+\frac{1}{2\sqrt{6}}R_{31}+\frac{1}{6\sqrt{2}}R_{32}+ \frac{1}{12}R_{33}\ ,\\
& d_3 = \frac{3}{4} -\frac{1}{2\sqrt{6}}R_{31}+\frac{1}{6\sqrt{2}}R_{32}+\frac{1}{12}R_{33}+ \frac{1}{\sqrt{3}}R_{21}-\frac{1}{3}R_{22}-\frac{1}{3\sqrt{2}}R_{23} \ ,\\
& d_4 = \frac 34 + \frac{1}{\sqrt{2}}R_{32}-\frac{3}{12}R_{33}\ .
\end{align*}

\section*{Acknowledgments}

We thank Andrzej Kossakowski, Jacek Jurkowski and Adam Rutkowski for discussions. As the paper was completed we learned 
from Seung-Hyeok Kye that the SPA conjecture is not true.

\end{document}